\pdfoutput=1
\documentclass{PoS}

\usepackage{amsmath}

\def\la{\langle}
\def\ra{\rangle}

\def\beq{\begin{equation}}
\def\eeq{\end{equation}}
\def\bea{\begin{eqnarray}}
\def\eea{\end{eqnarray}}
\def\barr{\begin{array}}
\def\earr{\end{array}}

\title{Investigation of the 1+1 dimensional Thirring model using the method of matrix product states}

\ShortTitle{Massive Thirring model from MPS}

\author{Mari Carmen Ba\~{n}uls\\
        Max Planck Institut f\"{u}r Quantenoptik, Garching 86748, Germany\\
        E-mail: \email{banulsm@mpq.mpg.de}}

\author{Krzysztof Cichy\\
        Faculty of Physics, Adam Mickiewicz University,
        ul. Umultowska 85, 61-614 Pozna\'{n},
        Poland\\
        E-mail: \email{krzysztof.cichy@gmail.com}}

\author{Ying-Jer Kao\\
        Department of Physics, National Taiwan University, Taipei 10617
        Taiwan\\
        E-mail: \email{yjkao@phys.ntu.edu.tw}}

\author{\speaker{C.-J.~David~Lin}\\
        Institute of Physics, National Chiao-Tung University,
        Hsinchu 30010, Taiwan\\
        Centre for High Energy Physics, Chung-Yuan Christian
        University, Chung-Li, 32032, Taiwan
        E-mail: \email{dlin@mail.nctu.edu.tw}}

\author{Yu-Ping Lin\\
        Department of Physics, University of Colorado,
        Boulder, CO 80309, USA\\
        E-mail: \email{Yuping.Lin@colorado.edu}}

\author{David T.-L Tan\\
        Institute of Physics, National Chiao-Tung University,
        Hsinchu 30010, Taiwan\\
        E-mail: \email{tanlin2013.py04g@nctu.edu.tw}}

\abstract{We present preliminary results of a study on the non-thermal
phase structure of the (1+1) dimensional massive Thirring model, employing the method of
matrix product states. Through investigating the entanglement
entropy, the fermion correlators and the chiral condensate, it is found that this approach enables us to
observe numerical evidence of a Kosterlitz-Thouless phase transition
in the model.}

\FullConference{The 36th Annual International Symposium on Lattice Field Theory - LATTICE2018\\
		22-28 July, 2018\\
		Michigan State University, East Lansing, Michigan, USA.}

\begin{document}
\section{Introduction}
\label{sec:introduction}
For the past few years, tensor-network (TN) methods for quantum many-body systems have been
attracting attention from practitioners of 
lattice field theory.   These methods feature several appealing
qualities, such as the opportunity for studying real-time dynamics of
quantum field theories (QFT's), and the
possibility of solving the sign problem.  In addition to exploring the
formulation of various
QFT's as TN's, there have also been efforts on applying
existing techniques to lattice field theory computations.  See. {\it
  e.g.}, Refs.~\cite{Banuls:2018latt, Banuls:2016jws} for a summary of recent activities. 

In this article, we present progress of our investigation for the non-thermal
phase structure of (1+1)-dimensional massive Thirring model using the formulation of
matrix product states (MPS).   The action of this model is
\begin{equation} 
\label{eq-action-thirring}
    S_{\mathrm{Th}}[\psi,\bar{\psi}] \
    = \int d^2x \left[ \bar{\psi}i \gamma^{\mu}\partial_{\mu}\psi \
      - m\,\bar{\psi}\psi \
      -\frac{g}{2} \left( \bar{\psi}\gamma_{\mu}\psi \right)\left( \bar{\psi}\gamma^{\mu}\psi \right) \right] \,,
\end{equation}
where $m$ and $g$ denote the fermion mass and the dimensionless
four-fermion coupling constant.   As shown by Coleman~\cite{Coleman:1974bu} and
Mandelstam~\cite{Mandelstam:1975hb}, the model in
Eq.~(\ref{eq-action-thirring}) is S-dual to the sine-Gordon (SG)
theory in the sector of zero total fermion number.
The SG theory is also
known to be dual to the classical two-dimensional XY
model~\cite{Jose:1976wc}\footnote{In addition to establishing the
  equivalence between partition functions of the SC theory and the
  two-dimensional Coulomb gas, it was also shown
that the XY-model vortices can be manifested as SG solitons~\cite{Jose:1976wc, Huang:1990via}.}.  Therefore, a Kosterlitz-Thouless (KT)
phase transition~\cite{Kosterlitz:1973xp} is expected in the Thirring model.  
Hints for this phase structure are also observed in the Thirring-model renormalisation group (RG) equations up
to ${\mathcal{O}}(m^{3})$ in perturbation theory~\cite{ZinnJustin:2000dr},
\beq
\label{eq:RGE_Thirring}
 \beta_{g} \equiv \mu \frac{d g}{d \mu} = -64 \pi
 \frac{m^{2}}{\Lambda^{2}} \mbox{ }, \mbox{ }\mbox{ }
 \beta_{m} \equiv \mu \frac{d m}{d \mu} = \frac{-2 (g +
   \frac{\pi}{2})}{g + \pi} m - \frac{256 \pi^{3}}{(g+\pi)^{2}
   \Lambda^{2}} m^{3} ,
\eeq
which bear the same characteristics as the Kosterlitz scaling
equations for the XY model~\cite{Kosterlitz:1974sm}.   For
$g < -\pi/2$, RG transformation
evolves the theory to the massless Thirring model which is a
conformal field theory (CFT).  In this regime, the
dual SG model is shown to be a free bosonic QFT at low energy.
At $g > -\pi/2$, the $\bar{\psi}\psi$ operator in
Eq.~(\ref{eq-action-thirring}) is relevant and the theory contains a scale.
 
In the current project, we demonstrate that the MPS formulation
can be employed to perform numerical lattice calculations for
investigating the phase structure of the massive Thirring model.
This is the foundation
work for our future study of real-time dynamics of the above phase
transition.   Exploratory results of this project were already
presented in Ref.~\cite{Banuls:2017evv}.

\section{Massive Thirring model and the XXZ quantum spin chain}
\label{sec:formulation}
Equation~(\ref{eq-action-thirring}) describes the classical action of
the Thirring model.
To obtain the operator formalism for this model as
a QFT, it is necessary to incorporate effects of the anomalous breaking of the
vector symmetry in two dimensions~\cite{Schwinger:1962tp}.  In this
procedure, the construction of the Hamiltonian can be
achieved by studying the operator relations that are satisfied by the
energy-momentum tensor~\cite{hagen1967new}.   Such relations result in
the continuum Thirring-model Hamiltonian operator,
\begin{equation} 
\label{eq-H_Th-continuum}
    H_{{\mathrm{Th}}} = \int dx \left[ -i Z_{\psi} (g)\bar{\psi}\gamma^{1}\partial_{1}\psi
    + m_{0}\bar{\psi}\psi
    + \frac{g}{4} \left(\bar{\psi}\gamma^{0}\psi\right)^2 
    - \frac{\tilde{g}(g)}{2} \left(\bar{\psi}\gamma^{1}\psi\right)^2
  \right] \, , 
\end{equation}
where $m_{0}$ is the bare mass, $Z_{\psi}(g)$ is the wavefunction
renormalisation constant~\cite{Mueller:1972md,Gomes:1972yb}, and $\tilde{g}(g) = g (g
+ \pi)/(2g + \pi)$.

Since we work in the Hamiltonian formalism, discretisation of the
theory only has to be carried out in the spatial direction. This is
achieved using staggered fermions~\cite{Banks:1975gq,
  Susskind:1976jm}.  In this strategy,  one of the two
components, $\psi_{1}$ and $\psi_{2}$, of the Dirac fermion is removed
on each lattice site.   This is implemented {\it via}
$\psi_{1}(x) \rightarrow \frac{1}{\sqrt{a}}\,c_{2n} \,  , \, \mbox{ }
    \psi_{2}(x) \rightarrow \frac{1}{\sqrt{a}}\,c_{2n+1} \,  , $
where $a$ is the lattice spacing, and $c_{i}$ is a one-component
fermionic degree of freedom at the $i{-}$th site of the
one-dimensional spatial lattice.  Choosing the standard representation of the Dirac matrices,   
$\gamma^{0} = \sigma^{z}, \mbox{ } \gamma^{1} = i \sigma^{y}, \mbox{ }
 \gamma^{5} = \sigma^{x}$ ($\{ \sigma^{i} \}$ are the Pauli matrices),   
%
we obtain the staggered Thirring-model Hamiltonian operator on the lattice,
\begin{equation} 
\label{eq-H_latt}
    H_{{\mathrm{Th}}}^{{\mathrm{(latt)}}} 
        = -\frac{i}{2a} Z_{\psi}(g) \sum_{n=0}^{N-2} 
            \Big( c^{\dagger}_{n}c_{n+1} - c^{\dagger}_{n+1}c_{n} \Big)
            + m_{0}\sum_{n=0}^{N-1} \left(-1\right)^{n} c^{\dagger}_{n}c_{n} 
        + \frac{\tilde{g}(g)}{2a} \sum_{n=0}^{N-2} \
            c^{\dagger}_{n}c_{n} c^{\dagger}_{n+1}c_{n+1} \, , 
\end{equation}
where $N$ is the total number of lattice sites.    Since we are
only discretising the spatial direction, the fermion doubling problem
is completely evaded in staggered fermions.   Therefore,
$H_{{\mathrm{Th}}}^{{\mathrm{(latt)}}}$ is describing one ``flavour'' of
fermion at the effective lattice spacing $2a$.

It is challenging to perform simulations with the fermionic
degrees of freedom in Eq.~(\ref{eq-H_latt}).  To proceed, we 
employ the Jordan-Wigner (JW) transformation to represent the
model using spin matrices.  This results
in a one-dimensional quantum XXZ spin chain coupled to uniform and staggered external
magnetic fields.  The system is described by the Hamiltonian~\cite{Luther:1976mt},
\bea
\label{eq:H_XXZ__to_sim}
 &&H_{{\mathrm{XXZ}}} = \frac{\nu (g)}{a} \bar{H}_{{\mathrm{sim}}} \,
 , \mbox{ }\mbox{ }
{\mathrm{with}}\mbox{ }\mbox{ }
 \bar{H}_{{\mathrm{sim}}} = -\frac{1}{2} \sum_{n=0}^{N-2} \left( S_{n}^{+}S_{n+1}^{-} 
            + S_{n+1}^{+}S_{n}^{-} \right)+ a \tilde{m}_{0} \sum_{n=0}^{N-1} \left(-1\right)^{n} \left(
              S_{n}^{z}+\frac{1}{2} \right)  \nonumber\\
           && \mbox{ }\mbox{ }\mbox{ }\mbox{ }\mbox{ }\mbox{ }\mbox{
           }\mbox{ }\mbox{ }\mbox{ }\mbox{ }\mbox{ }\mbox{ }\mbox{
           }\mbox{ }\mbox{ }\mbox{ }\mbox{ }\mbox{ }\mbox{ }\mbox{
           }\mbox{ }\mbox{ }\mbox{ }\mbox{ }\mbox{ }\mbox{ }\mbox{ }\mbox{ }\mbox{ }\mbox{ }\mbox{ }\mbox{ }\mbox{ }\mbox{ }\mbox{ }\mbox{ }\mbox{ }\mbox{ }\mbox{ }\mbox{ }\mbox{ }\mbox{ }\mbox{ }\mbox{ }\mbox{ }\mbox{ }\mbox{ } \mbox{ }\mbox{ }\mbox{ }\mbox{ }\mbox{ }\mbox{ }\mbox{ }\mbox{ }\mbox{ }\mbox{ }\mbox{ }\mbox{ }\mbox{ }\mbox{ }\mbox{ }\mbox{ }\mbox{ }\mbox{ }\mbox{ }\mbox{ }\mbox{ }\mbox{ }\mbox{ }\mbox{ }\mbox{ }\mbox{ }\mbox{ }\mbox{ }\mbox{ }\mbox{ }\mbox{ }\mbox{ }
        + \Delta (g) \sum_{n=0}^{N-1} \left( S_{n}^{z}+\frac{1}{2} \right) \ 
            \left( S_{n+1}^{z}+\frac{1}{2} \right) \,,
\eea
where $S_{n}^{\pm} = S_{n}^{x} \pm i S_{n}^{y}$ is the spin matrix ($S^{i}_{n} = \sigma^{i}/2$)
at the $n{-}$the lattice site, and $[S^{i}_{n}, S^{j}_{m}] = 0$ when
$n \not= m$.  The functions $\nu (g)$ and $\Delta (g)$ are the lattice
counterparts of $Z_{\psi} (g)$ and $\tilde{g} (g)$ in
Eq.~(\ref{eq-H_Th-continuum})~\cite{Luther:1976mt}, 
\beq
\label{eq:nu_and_Delta}
 \nu (g) = \left ( \frac{\pi - g}{\pi} \right )/ \sin\left (
 \frac{\pi - g}{2} \right ) \, , \mbox{ }\mbox{ }
   \Delta (g) = \cos \left (
 \frac{\pi - g}{2} \right )  \, , 
\eeq
and $\tilde{m}_{0} = m_{0}/\nu (g)$ is the rescaled bare mass.

%
\section{Matrix product states and simulation details}
\label{sec:MPS}
We continue the phase-structure study with the ground
state search in the Hamiltonian obtained by modifying
$\bar{H}_{{\mathrm{sim}}}$ in Eq.~(\ref{eq:H_XXZ__to_sim}),
\beq
\label{eq-penalty-term}
    \bar{H}_{{\mathrm{sim}}}^{{\mathrm{penalty}}} = \bar{H}_{{\mathrm{sim}}} + \lambda \left( \sum_{n=0}^{N-1} S_{n}^{z} \right)^2 \,,
\eeq
where $\lambda$ is a parameter chosen to be 100 in this work.  The
``penalty term'' guarantees that $\sum_{n=0}^{N-1} S_{n}^{z}
= 0$  for the ground state determined with a
variational search strategy~\cite{Banuls:2013jaa}.  Employing the JW transformation, it can
be demonstrated that this corresponds to constraining the Thirring
model in the sector of zero
total fermion number, in which the duality with the SG theory is valid.

The size of the Hilbert space in a quantum spin chain increases exponentially
with the system size, $N$.  This makes numerical
computations very challenging in practice.
The MPS method is an efficient, entanglement-based, way to truncate
the Hilbert space, and can be combined with the approach {\it \'{a}la}
density matrix RG (DMRG)~\cite{White:1992zz,Schollwock:2011ap} for
obtaining the ground state.   
Its implementation for representing a quantum state of $N$ spins ($\sigma_{1},\ldots,\sigma_{N}$) in one dimension can be
summarised as
\beq
\label{eq:MPS}
    |\Psi\rangle = \sum_{\sigma_{1},\ldots,\sigma_{N}}
    c_{\sigma_{1}\ldots\sigma_{N}}\,|\sigma_{1}\ldots\sigma_{N}\rangle
    \xrightarrow{\text{MPS}}\sum_{\sigma_{1},\ldots,\sigma_{N}} M^{\sigma_{1}}M^{\sigma_{2}} \ldots M^{\sigma_{N-1}}M^{\sigma_{N}}\,|\sigma_{1}\ldots\sigma_{N}\rangle \,,
\eeq
where each $M^{\sigma_{n}}$ is a matrix with maximally-allowed dimension (the bond
dimension) being $D$.  
The strategy in (3.2) involves an approximation of the rank${-}N$ tensor, 
$c_{\sigma_{1}\ldots\sigma_{N}}$, by $N$ rank${-}3$ tensors,
$\{ M^{\sigma_{n}}_{i,j} \}$.  This implies a truncation of the Hilbert space
which can be achieved {\it via} singular value (SV) decomposition with
respect to each cut of the chain, in which only the largest $D$ SV's 
are kept.  This programme often allows one to capture
low-energy physics with small enough $D$ in practice.

To carry out the DMRG search for the ground state of
$\bar{H}_{{\mathrm{sim}}}^{{\mathrm{penalty}}}$, we also need the
matrix product operator for this Hamiltonian.   Through explicit
construction, we find 
[$\beta_{n} \equiv \Delta (g) + (-1)^{n} \tilde{m}_{0} a$],
\beq
\label{eq:H_MPO}
 \bar{H}_{{\mathrm{sim}}}^{{\mathrm{penalty}}} = 
 \prod_{n=0}^{N-1} W^{[n]}\, ,
\eeq
where
\bea
\label{eq:MPO_bound}
 && W^{[0]} = \left ( \begin{array}{cccccc}
   {\mathbb{I}}\mbox{ }\mbox{ } &  
   \mbox{ }\mbox{ }  -\frac{1}{2} S^{+}_{0} \mbox{ }\mbox{ } &
   \mbox{ }\mbox{ }   -\frac{1}{2} S^{-}_{0} \mbox{ }\mbox{ } & 
   \mbox{ }\mbox{ }  2\lambda  S^{z}_{0} \mbox{ }\mbox{ }  & 
   \mbox{ }\mbox{ }  \Delta (g) S^{z}_{0} \mbox{ }\mbox{ } &
   \mbox{ }\mbox{ } \beta_{0} S^{z}_{0} + \frac{\lambda}{4}\\
 \end{array}\right ) \, , \nonumber\\
 && W^{[N-1]} = \left ( \begin{array}{cccccc}
    \beta_{N-1} S^{z}_{N-1} + \frac{\lambda}{4}\mbox{ }\mbox{ } &  
   \mbox{ }\mbox{ }  S^{-}_{N-1} \mbox{ }\mbox{ } &
   \mbox{ }\mbox{ }  S^{+}_{N-1} \mbox{ }\mbox{ } & 
   \mbox{ }\mbox{ }  S^{z}_{N-1} \mbox{ }\mbox{ }  & 
   \mbox{ }\mbox{ }  S^{z}_{N-1} \mbox{ }\mbox{ } &
   \mbox{ }\mbox{ } {\mathbb{I}}\\
 \end{array}\right )^{T} \, ,
\eea
and $W^{[n]}|_{2\le n \le N-2}  \equiv W^{[n]}_{{\mathrm{bulk}}}$ is
a $6\times 6$ operator-valued matrix,
\bea
\label{eq:MPO}
&& \left ( W^{[n]}_{{\mathrm{bulk}}} \right )_{1j} = \left (
  W^{[0]}|_{S^{+,-,z}_{0}, \beta_{0} \rightarrow S^{+,-,z}_{n}, 
    \beta_{n}} \right )_{1j} \, , \mbox{ } \left ( W^{[n]}_{{\mathrm{bulk}}} \right )_{i6} = \left (
  W^{[N-1]}|_{S^{+,-,z}_{N-1}, \beta_{N-1} \rightarrow S^{+,-,z}_{n}, 
    \beta_{n}} \right )_{i1} \, ,\nonumber\\
&& \left ( W^{[n]}_{{\mathrm{bulk}}} \right )_{44} = {\mathbb{I}} \, ,
\mbox{ }{\mathrm{and}}\mbox{ }{\mathrm{all}}\mbox{
}{\mathrm{other}}\mbox{ } \left ( W^{[n]}_{{\mathrm{bulk}}} \right
)_{ij} = 0 \, .
\eea

In this work, we perform simulations at four system sizes, $N = 400,
600, 800, 1000$.  This enables us to extrapolate our results to the
infinite-volume limit.  To carry out the initial scanning for studying the
phase structure, twenty-one values of the bare four-fermion
coupling are chosen, straddling the range $-0.9 \leq \Delta (g) \leq 1.0$,
and five values of $a\tilde{m}_{0}$ ($ 0, 0.1, 0.2,
0.3, 0.4$) are selected.   As explained in the next section,
simulations at smaller fermion masses are needed, and we have been
generating data at $a\tilde{m}_{0} = 0.005, 0.01, 0.02, 0.03, 0.04,
0.06, 0.08, 0.13, 0.16$.

\begin{figure}[t!]
\begin{center}
\vspace{-0.5cm}
        \includegraphics[width=6.0cm, height=4.5cm]{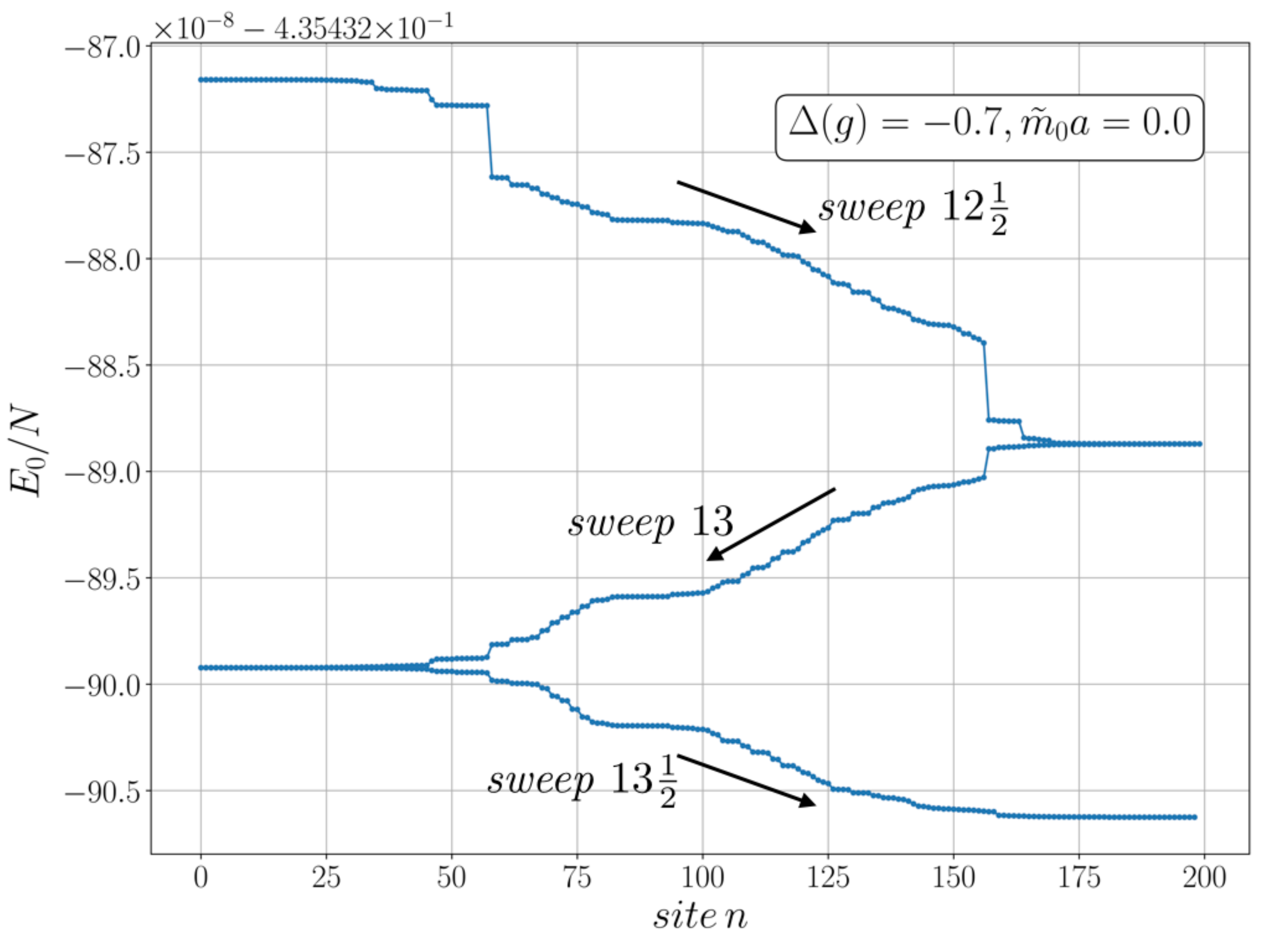}
   \hspace{0.5cm}
        \includegraphics[width=6.0cm,
        height=4.5cm]{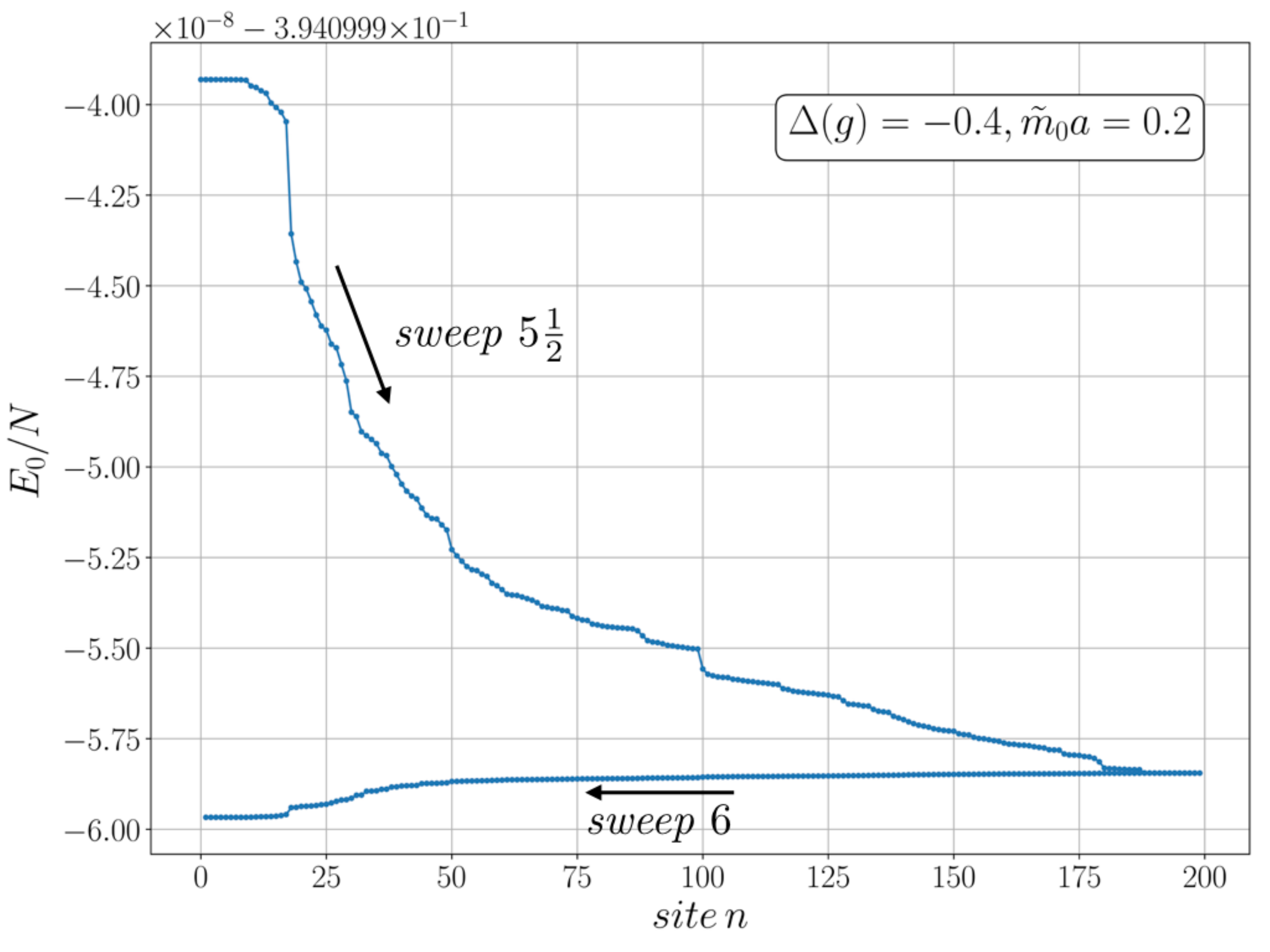}
\vspace{-0.2cm}
       \caption{Convergence of DMRG with two choices of
         $[\Delta(g),a\tilde{m}_{0}]$ at $N=200$ and $D=50$.}
\label{fig:DMRG_convergence}
\end{center}
\end{figure}

We commence the DMRG ground-state search at $D=50$, with
a set of random initial matrices, $\{ M^{\sigma_{n}} \}$.  
Results at six larger bond dimension, $D=100, 200, \ldots, 600$ are
then subsequently obtained by using the $\{ M^{\sigma_{n}} \}$ at smaller
$D$ as the initial input.  Having data at many values of the bond dimension enables
us to investigate the infinite${-}D$ limit.  The variational update of the MPS is performed
locally~\cite{Schollwock:2011ap} by sweeping over the lattice.  The target precision of the ground-state
energy is $10^{-8}$ in lattice units.  Figure~\ref{fig:DMRG_convergence} shows examples of how the DMRG
search converges at two choices of the bare parameters.   Notice that
significantly different convergence properties are observed in our
simulations, leading to hints that there are distinct phases in
the theory.
\section{Numerical results for the phase structure of the massive
  Thirring model}
\label{sec:phase_structure}
To investigate the non-thermal phase structure of the massive Thirring
model, we first examine the von Neumann entanglement entropy, computed
by dividing the system of size $N$ into two parts between sites $n$
and $n+1$.   Denoting this
entropy by $S_{N}(n)$, a scaling law is valid when the theory is conformal~\cite{Calabrese:2004eu}, 
\beq
\label{eq:Calabrese_Cardy}
 S_{N}(n) = \frac{c}{6} {\mathrm{ln}} \left [ \frac{N}{\pi}
   {\mathrm{sin}} \left ( \frac{\pi n}{N} \right ) \right ] + k \mbox{
   }\mbox{ }\left ( n = 0, 1, 2,\ldots , N-1\right ) \, , 
\eeq
where $c$ is the central charge, and $k$ is a constant.
\begin{figure}[t!]
\begin{center}
\vspace{-0.5cm}
        \includegraphics[width=4.9cm, height=4.5cm]{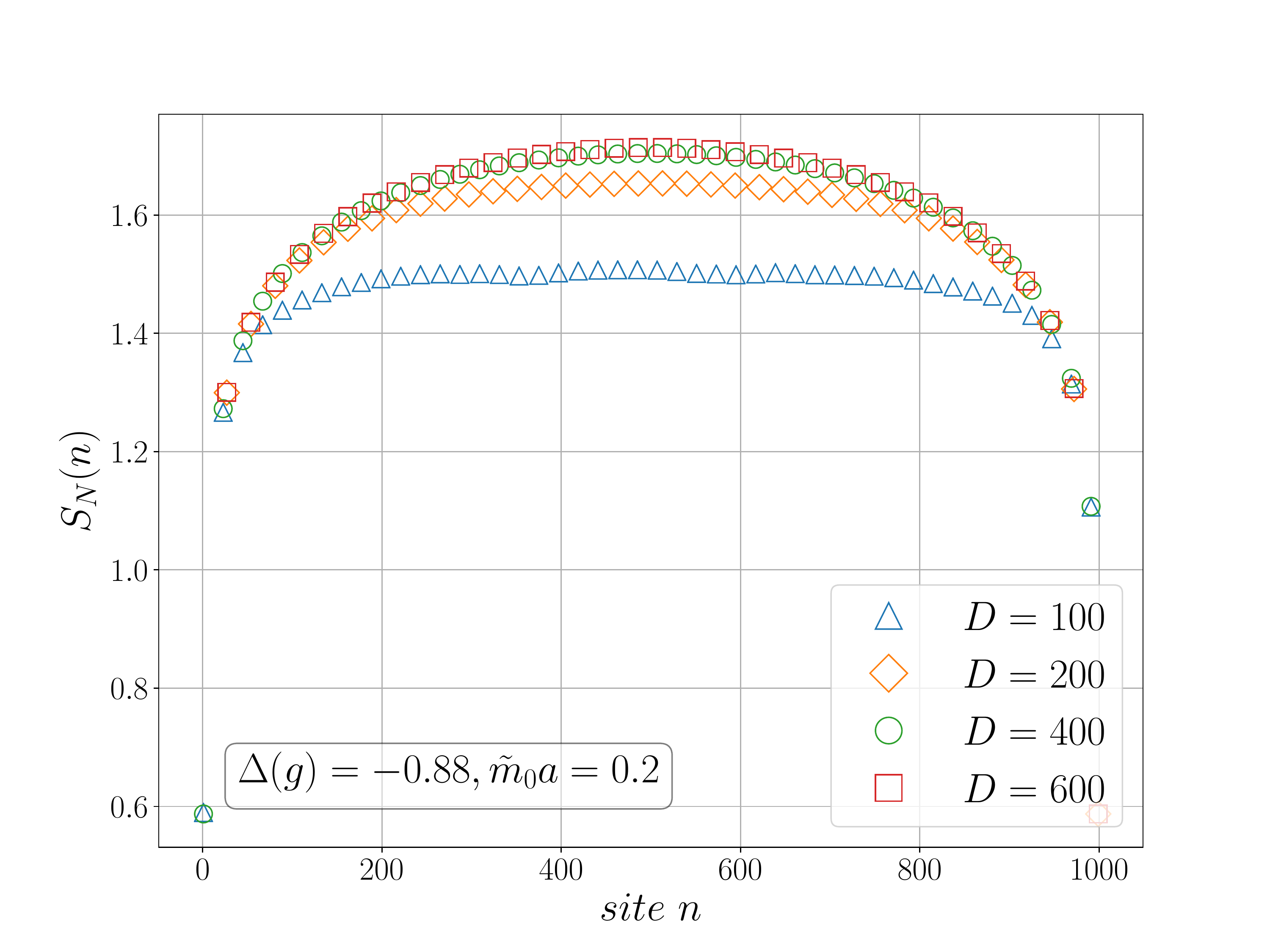}
        \includegraphics[width=4.9cm,
        height=4.5cm]{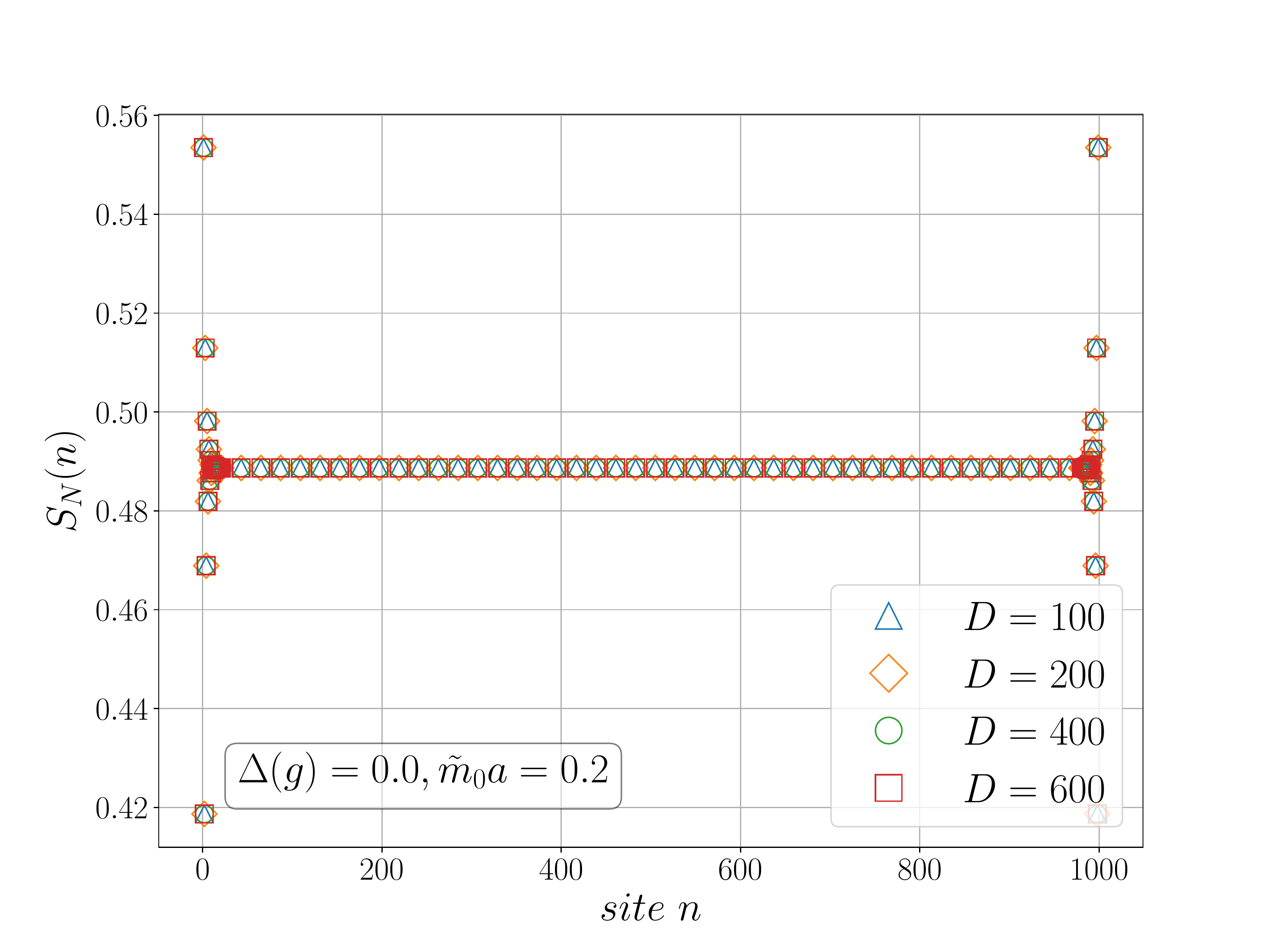}
        \includegraphics[width=4.9cm,
        height=4.5cm]{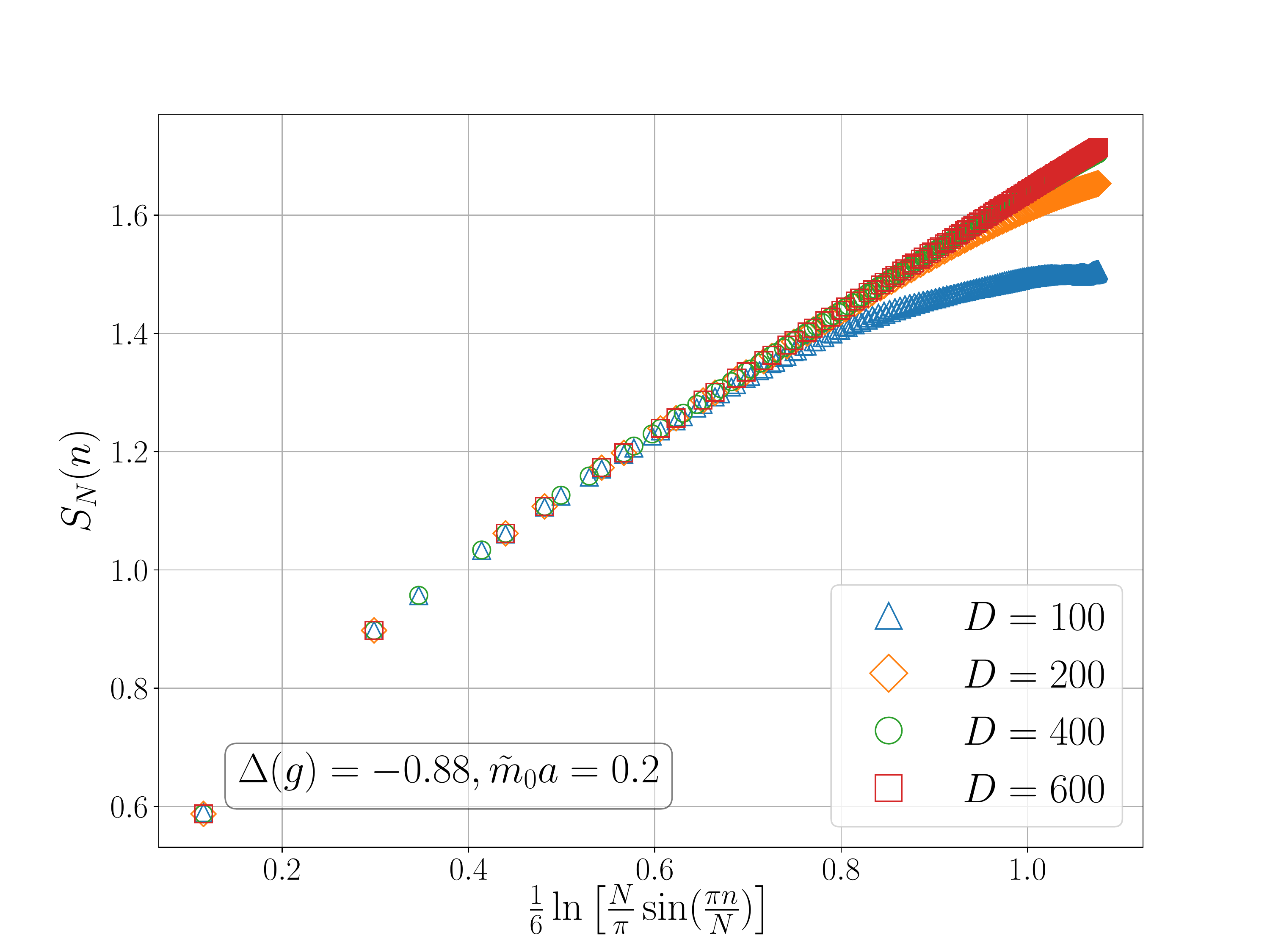}\\
        (a) \hspace{4.3cm} (b) \hspace{4.3cm} (c) 
      \vspace{-0.1cm}
       \caption{$S_{N}(n)$ at $N=1000$ plotted against (a) and (b): $n$ and (c): $\frac{1}{6} {\mathrm{ln}} \left [ \frac{N}{\pi}
   {\mathrm{sin}} \left ( \frac{\pi n}{N} \right ) \right ]$.}
\label{fig:EE_CC_scaling}
\end{center}
\end{figure}
Figure~\ref{fig:EE_CC_scaling}(a) demonstrates that this scaling
is manifested at $a\tilde{m}_{0}=0.2$ and $\Delta (g) = -0.88$, with
large enough $D$, and Fig.~\ref{fig:EE_CC_scaling}(c) shows that $c=1$
in this case, indicating that the theory in this regime is equivalent to a
free bosonic QFT.
For the same choice of $a\tilde{m}_{0}=0.2$, the behaviour in
Eq.~(\ref{eq:Calabrese_Cardy}) is not realised when $\Delta (g)$ is
increased beyond a critical value.  This is evidenced by the plot in
Fig.~\ref{fig:EE_CC_scaling}(b).  Such a feature persists in all our
results at $a\tilde{m}_{0}\not= 0$, while we find that the theory is
always conformal with $c=1$ when $a\tilde{m}_{0}=0$.   

\begin{figure}[t!]
\begin{center}
\vspace{-0.5cm}
        \includegraphics[width=6.0cm, height=4.5cm]{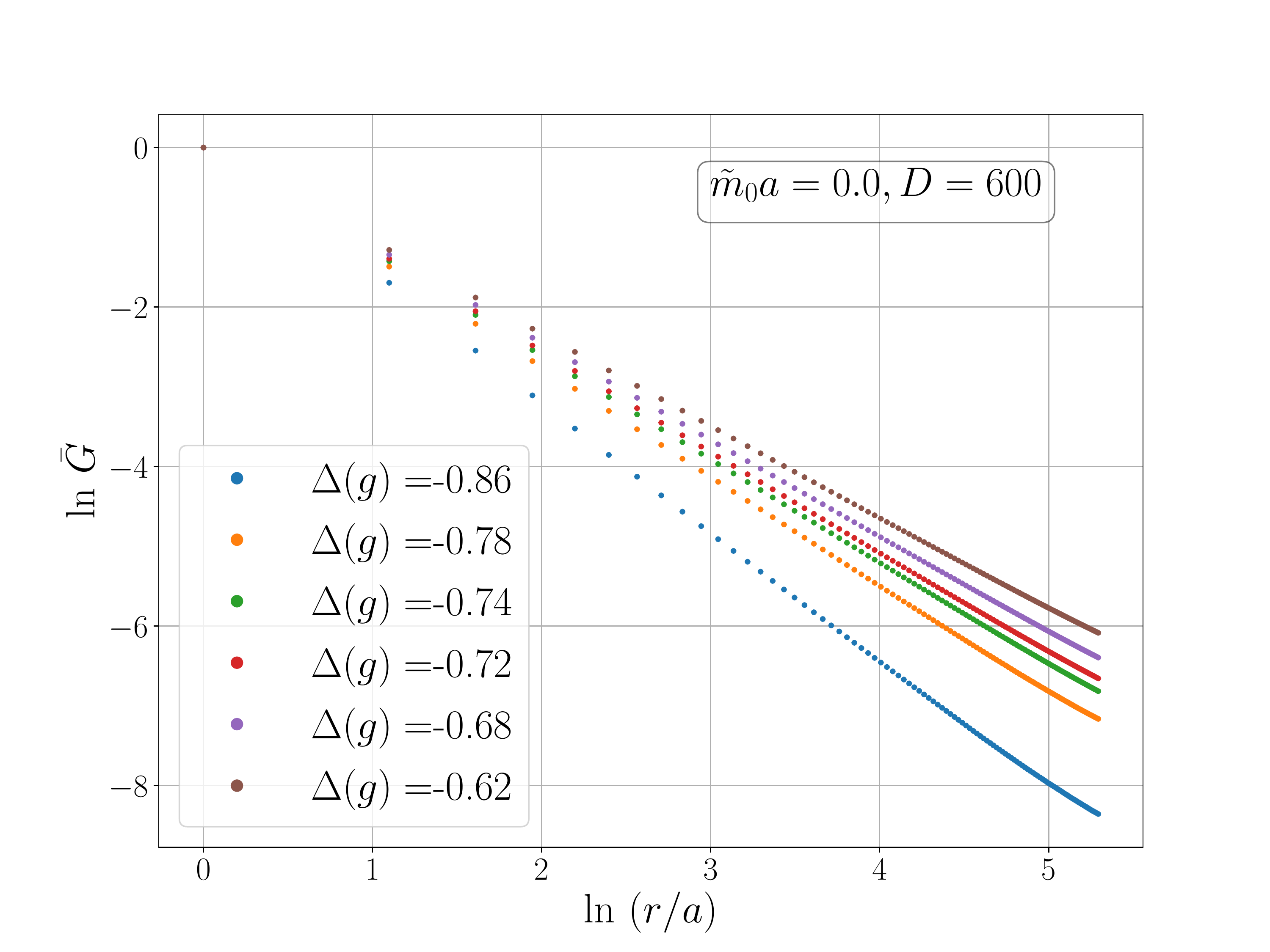}
   \hspace{0.5cm}
        \includegraphics[width=6.0cm,
        height=4.5cm]{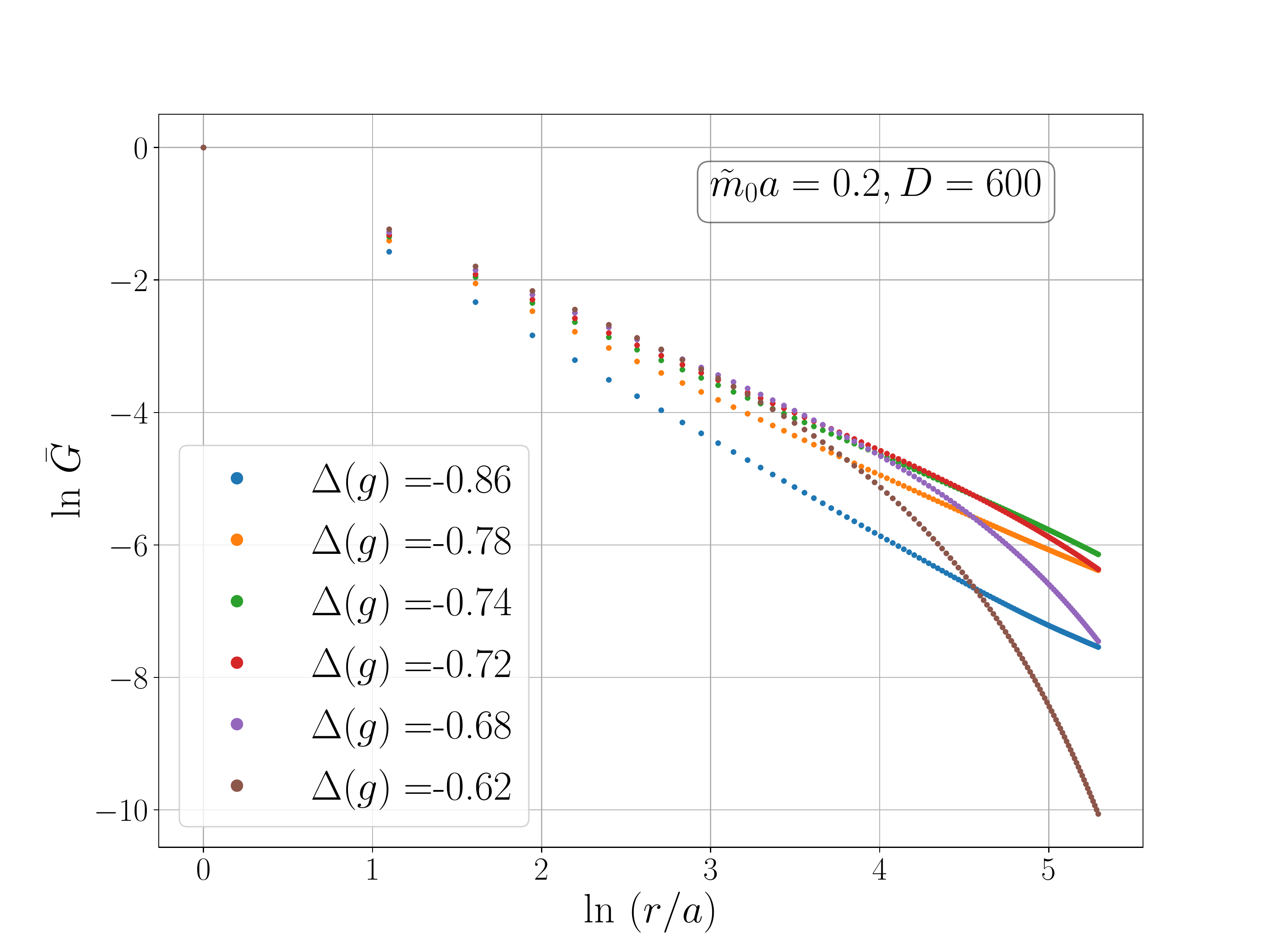}
\vspace{-0.2cm}
       \caption{Fermion (soliton) correlators at $a\tilde{m}_{0} =
         0.0$ and 0.2.}
\label{fig:string_corr}
\end{center}
\end{figure}

Next, we study the correlator of left-handed fermion fields,
$\psi_{{\mathrm{L}}}$(0) and $\psi_{{\mathrm{L}}}^{\dagger}(\hat{r})$, with $\hat{r} \equiv r/a$,
\beq
\label{eq:fermion_corr}
 \bar{G}(\hat{r}) \equiv G(\hat{r})/G(0) \, , \mbox{ }\mbox{ }{\mathrm{where}}
 \mbox{ } G(\hat{r}) = \la \psi_{{\mathrm{L}}}^{\dagger} (\hat{r})
 \psi_{{\mathrm{L}}} (0) \ra \xrightarrow{{\mathrm{JW}}\mbox{
   }{\mathrm{transformation}}}  \la S^{+}_{\hat{r}}   {\mathrm{e}}^{ i\pi
   \sum_{j=1}^{\hat{r}-1}  S^{z}_{j}  }   S^{-}_{0} \ra \, .
\eeq
This correlator is bosonised to the soliton correlator in the SG
model~\cite{Mandelstam:1975hb},  and exhibits power law in $\hat{r}$
in the conformal phase, while decays exponentially when the theory is
gapped~\cite{Witten:1978qu}.  From 
Fig.~\ref{fig:string_corr}, we see evidence that the model is
conformal when $a\tilde{m}_{0}=0$.  For $a\tilde{m}_{0}=0.2$ (also observed for
other values of non-zero $a\tilde{m}_{0}$), there
is a phase transition.  
Furthermore, the exponent for the power-law decay of this correlator
depends on $g$.  This leads to hints that the transition is of KT-type.

\begin{figure}[t!]
\begin{center}
\vspace{-0.5cm}
   \hspace{-0.7cm}
        \includegraphics[width=6.0cm, height=4.5cm]{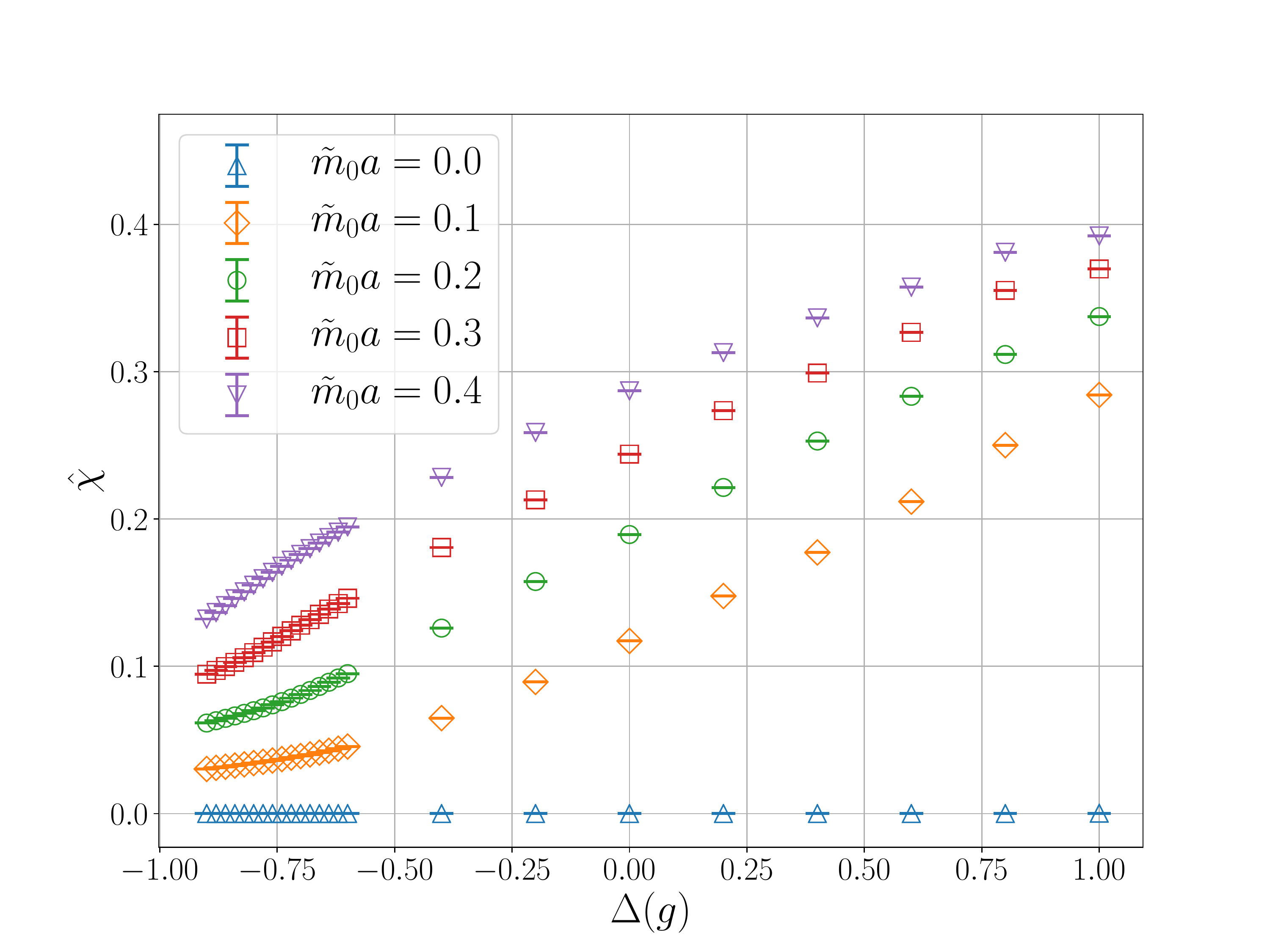}
   \hspace{0.7cm}
        \includegraphics[width=5.8cm,
        height=4.0cm]{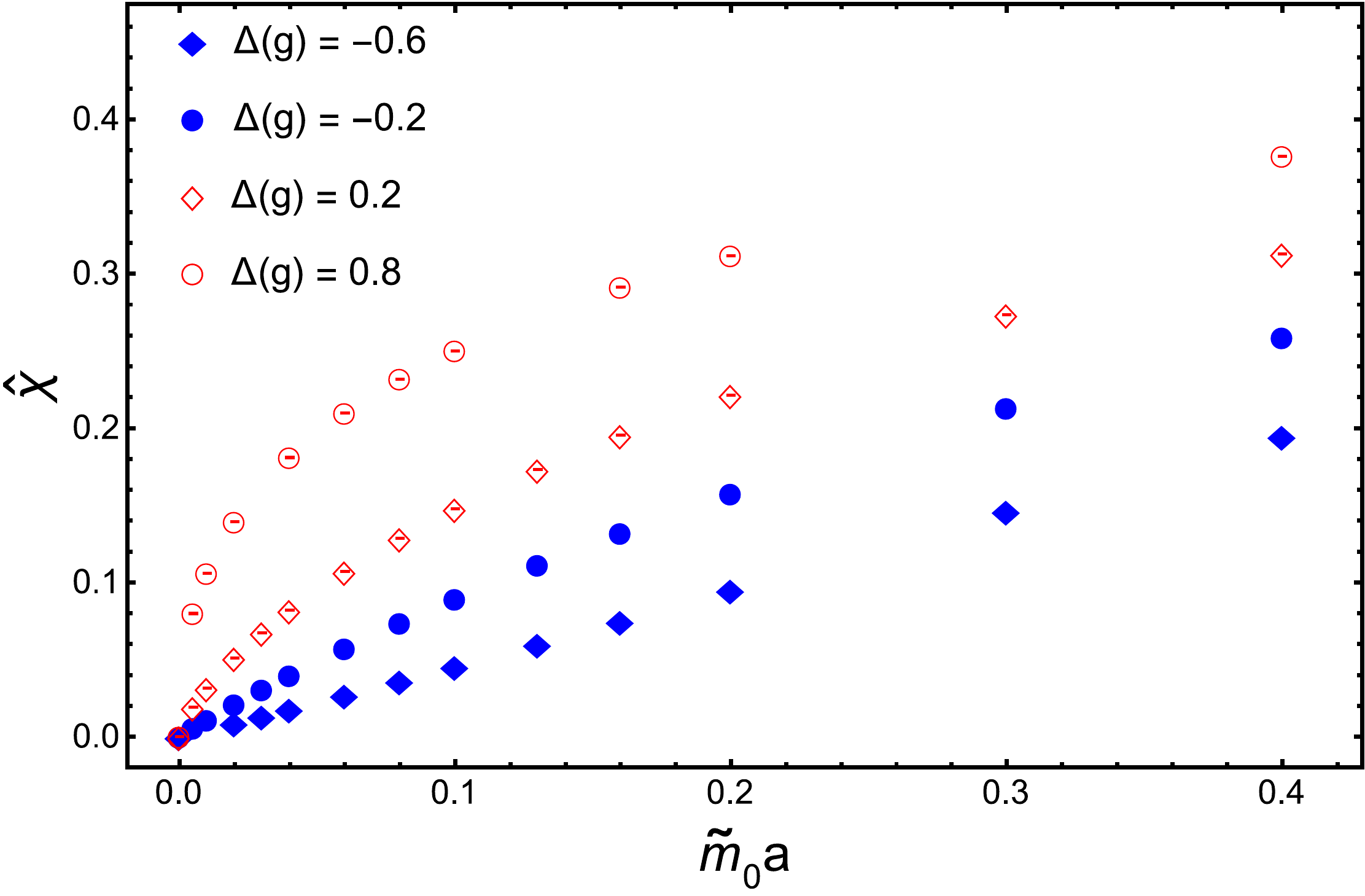}\\
  \hspace{-0.2cm} (a)\hspace{6.5cm} (b)
\vspace{-0.2cm}
       \caption{Results of chiral condensate in the limit of infinite
         bond dimension and system size.}
\label{fig:chiral_cond}
\end{center}
\end{figure}

The last quantity we investigate is the chiral condensate, $\hat{\chi}$.  Under the
JW transformation,
\beq
\label{eq:condensate}
 \hat{\chi} \equiv \frac{a}{N} \sum_{n=0}^{N-1} \la \bar{\psi} (n)\psi
 (n) \ra \xrightarrow{{\mathrm{JW}}\mbox{
   }{\mathrm{transformation}}}  \frac{1}{N} \la \sum_{n=0}^{N-1}
 (-1)^{n} S^{z}_{n} \ra .
\eeq
Results for this quantity at $a\tilde{m}_{0} = 0, 0.1, 0.2, 0.3$ and
$0.4$, extrapolated to infinite $D$ and $N$,  are displayed in
Fig.~\ref{fig:chiral_cond}(a).  We also performed direct
infinite-size simulations for $a\tilde{m}_{0} = 0$ using the uniform
MPS method~\cite{Zauner-Stauber:2018kxg}, and obtain $\hat{\chi}=0$.  To confirm that we are observing a KT 
transition, it is crucial to verify that the chiral condensate
computed with non-vanishing mass extrapolates to zero in the massless limit~\cite{Witten:1978qu}.
Figure~\ref{fig:chiral_cond}(b) shows this behaviour in both gapped
and conformal phases\footnote{ A vanishing condensate at $a\tilde{m}_{0} = 0$ is
consistent with the Mermin-Wagner-Coleman theorem~\cite{Mermin:1966fe,Coleman:1973ci}.}.  It also
demonstrates the importance of having results at $a\tilde{m}_{0} < 0.1$.
Currently we are performing more simulations in this regime.
\section{Conclusion and outlook}
\label{eq:conclusion}
In this article, we present our investigation of the non-thermal phase
structure of the (1+1) dimensional massive Thirring model, employing the MPS strategy.
We find that this approach can be successfully applied to such
studies, and numerical evidence for an expected KT phase transition is
observed.  To further confirm this observation, we are currently
examining the spectrum of the transfer matrix in this model.  Since the MPS method is formulated in the Hamiltonian
formalism, in the near future we will also use this technique to probe
real-time dynamics of this phase transition.

\section*{Acknowledgments}
MCB is supported by the QUANTERA project QTFLAG.  KC
was partially supported by the DFG project nr. CI 236/1-1. YJK,
CJDL, YPL and
DTLT acknowledge funding from Taiwanese MoST via grants
105-2628-M-009-003-MY4, 105-2112-M-002 -023 -MY3, 104-2112-M-002 -022
-MY3.

\end{document}